\documentclass{aastex}
\usepackage{spr-astr-addons}
\usepackage{url}
\RequirePackage{color}

\begin{document}

\title{A multiwavelength study of the supernova remnant G296.8$-$0.3}
\slugcomment{Not to appear in Nonlearned J., 45.}
\shorttitle{Short article title}
\shortauthors{Autors et al.}

\author{E. S\'anchez-Ayaso\altaffilmark{1}} \and \author{J.A. Combi\altaffilmark{2,5}}
\and
\author{J.F. Albacete Colombo\altaffilmark{3}} \and 
\author{J. L\'opez-Santiago\altaffilmark{4}} \and 
\author{J. Mart\'{\i}\altaffilmark{1}} \and 
\author{A.J. Mu\~noz-Arjonilla\altaffilmark{1}}



\altaffiltext{1} {Deptamento de F\'isica (EPS), Universidad de Ja\'en, Campus Las Lagunillas s/n Ed. A3 Ja\'en, Spain, 23071 email: [esayaso:jmarti:ajmunoz]@ujaen.es}
\altaffiltext{2} {Instituto Argentino de Radioastronom\'{\i}a (CCT La Plata, CONICET), C.C.5, (1894) Villa Elisa, Buenos Aires, Argentina. email: jcombi@fcaglp.unlp.edu.ar}
\altaffiltext{3} {Centro Universitario Regional Zona Atl\'antica (CURZA). Universidad Nacional del COMAHUE, Monse\~nor Esandi y Ayacucho (8500), 
Viedma (Rio Negro), Argentina. email: donfaca@gmail.com}
\altaffiltext{4} {Departamento de Astrof\'{\i}sica y Ciencias de la Atm\'osfera, Universidad Complutense de Madrid, E-28040, Madrid, Spain. email: jls@astrax.fis.ucm.es}   
\altaffiltext{5} {Facultad de Ciencias Astron\'omicas y Geof\'{\i}sicas, Universidad Nacional de La Plata, Paseo del Bosque, B1900FWA La Plata, Argentina.}

\begin{abstract}
We report XMM-{\it Newton} observations of the Galactic supernova remnant G296.8$-$0.3, together with complementary radio and infrared data. The spatial and spectral properties of the X-ray emission, detected towards G296.8$-$0.3, was investigated in order to explore the possible evolutionary scenarios and the physical connexion with its unusual morphology detected at radio frequencies. G296.8$-$0.3 displays diffuse X-ray emission correlated with the peculiar radio morphology detected in the interior of the remnant and with the shell-like radio structure observed to the northwest side of the object. The X-ray emission peaks in the soft/medium energy range (0.5-3.0 keV). The X-ray spectral analysis confirms that the column density is high ($N_{\rm H}$$\sim$0.64$\times$10$^{22}$ cm$^{-2}$) which supports a distant location (d$>$9 kpc) for the SNR. Its X-ray spectrum can be well represented by a thermal (PSHOCK) model,  with $kT$ $\sim$ 0.86 keV, an ionization timescale of 6.1$\times$10$^{10}$ cm$^{-3}$ s, and low abundance ($\sim$ 0.12 $Z_{\odot}$). The 24 $\mu$m observations show shell-like emission correlated with part of the northwest and southeast boundaries of the SNR. In addition a point-like X-ray source is also detected close to the geometrical center of the radio SNR. The object presents some characteristics of the so-called compact central objects (CCO). Its X-ray spectrum is consistent with those found at other CCOs and the value of $N_{\rm H}$ is consistent with that of G296.8$-$0.3, which suggests a physical connexion with the SNR.
\end{abstract}



\section{Introduction}

\begin{figure*}[t]
\includegraphics[width=8.5cm]{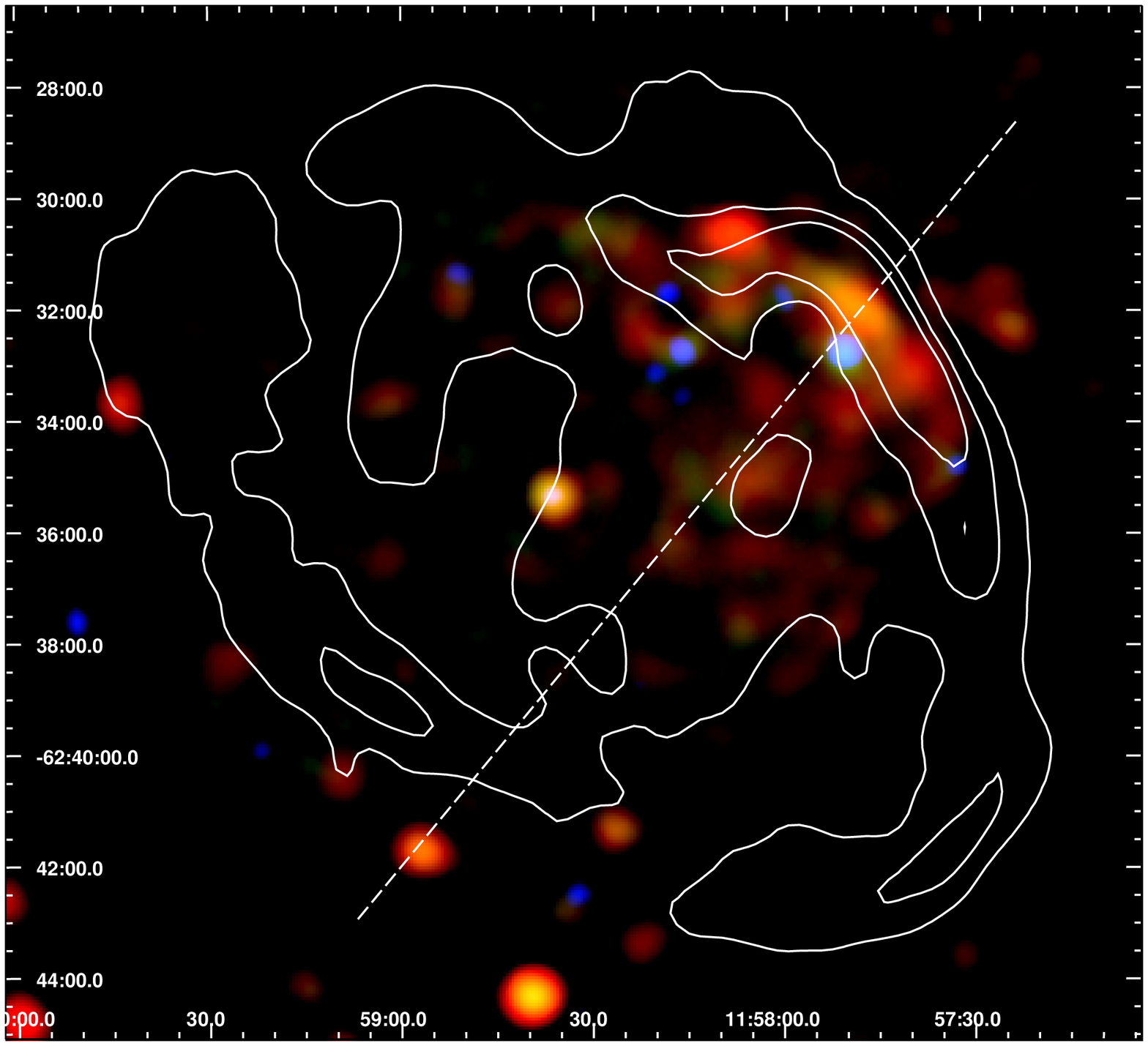}
\includegraphics[width=8.7cm]{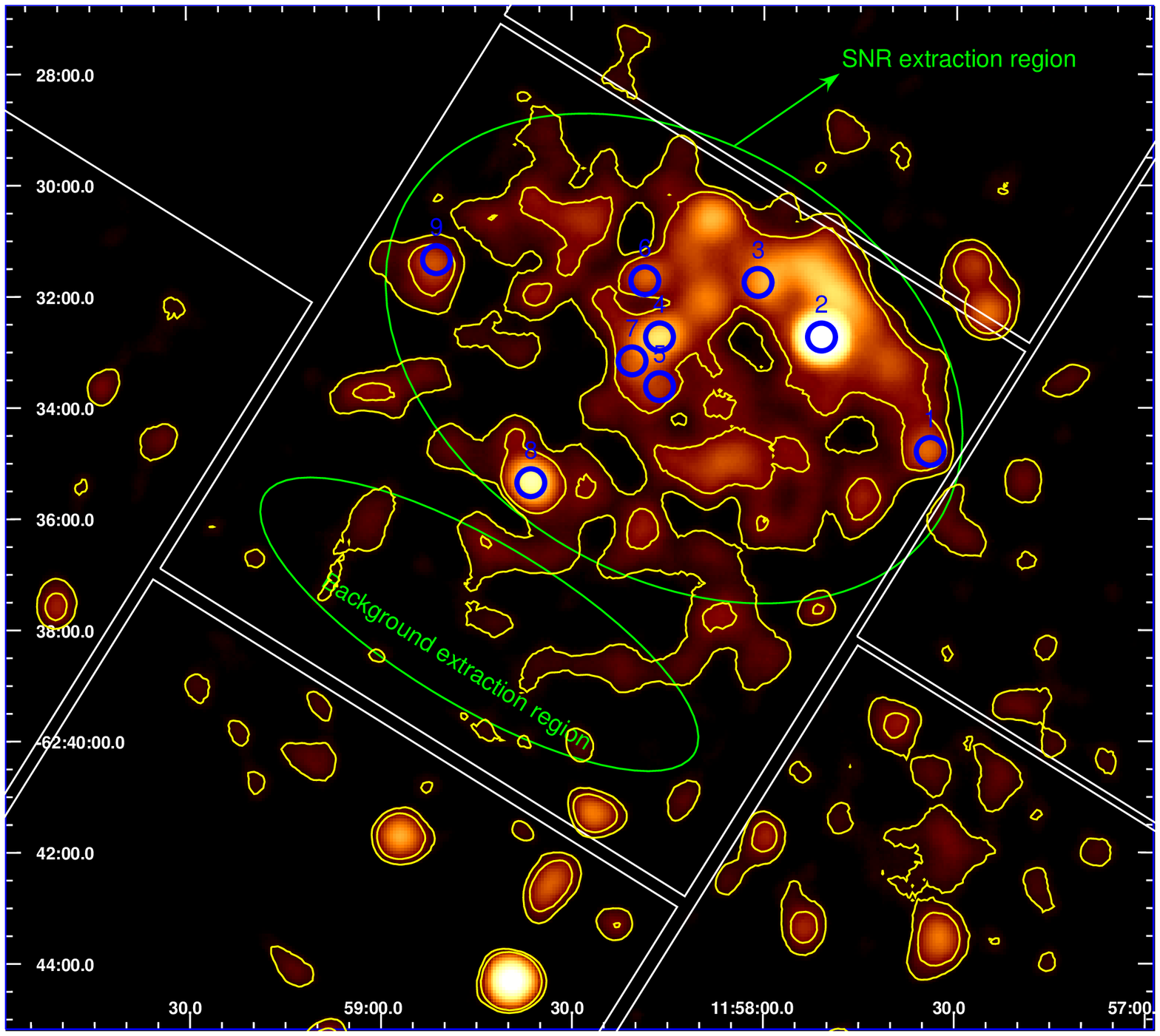}
   \caption{{\bf Left panel:} XMM-{\it Newton} X-ray color image of G296.8-0.3 generated for three energy bands: 0.5-1.0 (red), 1.0-2.0 (green), and 2.0-5.0 keV (blue), for the combined MOS1/2  cameras. A Gaussian smoothing with a kernel radius of 19.6$\times$19.6 arcsec was applied to the image. For comparison, the MOST radio image (in white contours) obtained at 843 MHz (Whiteoak \& Green 1996) is also presented in Figure. Radio contours are indicated in steps of 0.2, 0.29, and 0.83 mJy beam$^{-1}$. The dashed line indicates the direction of the low density tunnel in the interior of the remnant suggested by Gaensler et al. (1998). {\bf Right panel:} X-ray image of G296.8-0.3 in a logarithmic flux scale (units of ph cm$^{-2}$ s$^{-1}$) in the 0.5-5.0 keV energy range. The 1 and 3 sigma contours are showed in yellow color. The locations of the point-like X-ray sources detected within the SNR are numbered (in increasing right ascension order) and indicated by (blue) circles with radii of 15". The CCD size of the MOS camera and the extraction regions used for the spectral analysis are also indicated.}
    \end{figure*}

It is generally accepted that supernova remnants (SNRs) are the dominant source of Galactic cosmic rays,
at least for energies up to 3$\times$10$^{15}$ eV. These objects provide a significant fraction of the mechanical
energy that heats, compresses and chemically enriches the interstellar medium
(ISM). Therefore, SNRs can be used to investigate global properties of
the galaxy as well as the local environment where they evolve.
Thanks to significant advances in the angular resolution capabilities of modern X-ray observatories such as
XMM-{\it Newton} and {\it Chandra}, important progress has been made concerning the
detection of new and well-known SNRs (Sasaki et al. 2004; Bocchino et al. 2005; 
Combi et al. 2010a; Combi et al. 2010b). At present, 30\% of the radio SNRs display X-ray emission and more than a dozen of SNRs 
were originally discovered through their X-ray emission (e.g., Schwentker 1994;
Aschenbach 1998; Bamba et al. 2003; Yamaguchi et al. 2004). Studying the connexion 
between radio, infrared and X-ray emission of SNRs enables us to explore how stars end their lives and better understand the evolutionary process in this kind of fascinating sources.

The southern galactic SNR G296.8-0.3 (1156-62) lies in the direction to the Scutum-Crux arm of our galaxy. It was first detected 
at radio frequencies by Large \& Vaughan (1972). Initially, it was considered as a shell-type remnant 
despite its unusual and ill-defined shape (Shaver \& Goss 1970; Goss \& Shaver 1970). Subsequent higher-resolution radio observations 
at 843 MHz (Whiteoak \& Green 1996) showed a complicated multi-ringed structure, with its diffuse interior emission being brightest in its northwest side. HI observations carried out by Gaensler et al. (1997) allowed to obtain lower and upper limits on its systemic velocity in the range +15 to +30 km s$^{-1}$. This corresponds to a distance of 9.6$\pm$0.6 kpc. A flux density of 7.0$\pm$ 0.3 Jy was measured at 1.3 GHz for the total radio structure. Throughout this paper, a mean distance of 9 kpc is assumed. It corresponds to an angular size of $\sim$ 31 pc.

At X-ray energies, an exploratory study was carried out with the ROSAT satellite by Hwang \& Markert (1994). These authors noted marginal X-ray emission (4 $\sigma$ detection) near the peak of radio emission to the northwest part of the remnant. However, the poor statistics and limited X-ray energy range of the ROSAT telescope did not allow to observe a clear well-defined morphology of the X-ray emission.   

As part of a program addressed to study the X-ray emission of supernova remnants, in this paper we report XMM$-${\it Newton} observations of G296.8-0.3, and infrared data obtained with the Spitzer Space Telescope in order to study the physical characteristics of the object and the surrounding ISM where it evolves. The structure of the paper is as follows: in Sect. 2, we describe
the XMM-{\it Newton} observations, data reduction and present our X-ray analysis. IR results are reported in Sect. 3. In Sect. 4, we discuss the implications of our results. Finally in Sect. 5, we summarize our main conclusions.

\begin{figure*}[t]
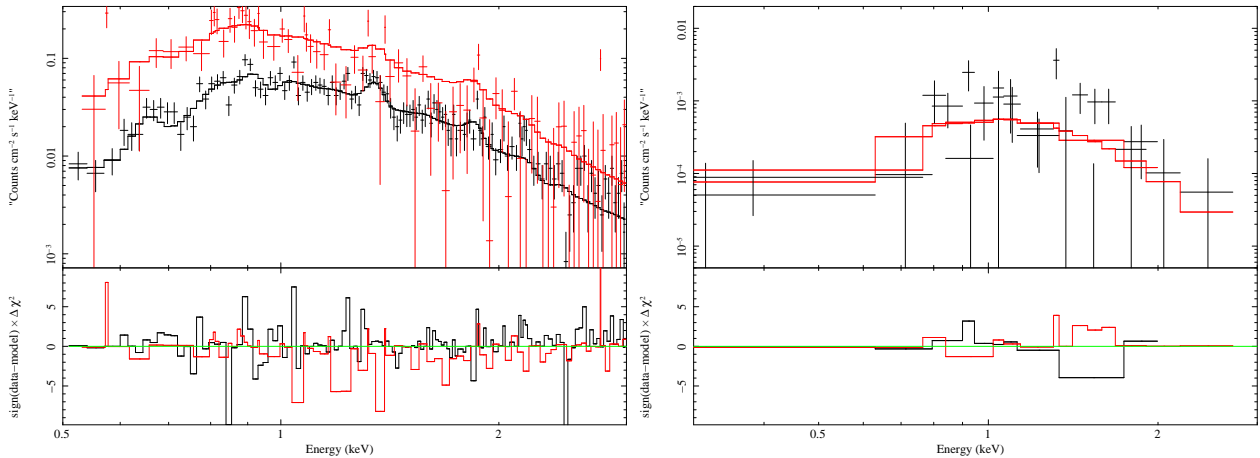

\includegraphics[width=6.0cm,angle=270]{fig-sa-2A.eps}
\includegraphics[width=6.0cm,angle=270]{fig-sa-2B.eps}
\caption{{\bf Left panel:} EPIC MOS1 (in black) and pn (in red) spectra of the SNR. Solid line indicate the best-fit model (see Table 1). {\bf Right panel:} EPIC MOS 1 and 2 spectra of the central point source. The bottom plots in the panels display the $\chi^2$ residual of the best fit model.}
\end{figure*}

\section{XMM-${\it Newton}$ observation and data reduction}

G296.8-0.3 was observed by the XMM-Newton X-ray satellite in two separate pointings. These were performed on 2008 February 16 (ObsID 0503780301) and 2008 August 16 (ObsID 0550170101), with the EPIC MOS (Turner et al. 2001) and EPIC pn (Str{\"u}der et al. 2001) cameras. Both observations have similar pointing coordinates ($\alpha$=11$^h$58$^m$30.0, $\delta$=$-$62$^{\circ}$35$^\prime$00.0 ; J2000), and were placed at the CCD center. The XMM-{\it Newton} data were calibrated and analyzed with the XMM Science Analysis System (SAS) version 10.0.0. To exclude high background activity, which can affect the observations, we extracted light curves of photons above 10 keV from the entire field-of-view of the cameras, and excluded time intervals with count rate higher than 3 $\sigma$ above average to produce a GTI file. Unfortunately, the observation ObsID 0503780301 was affected by a high and variable soft proton background level (Lumb et al. 2002), whereas the other one ObsID 0550170101 is unaffected by background fluctuations. In order to avoid contamination for high background patterns, hereafter our analysis concerns only the second observation. 

After the time filtering, 45.2 ks of useful data for MOS1, 45.1 ks for MOS2, and 46.5 ks for the pn cameras are available for further data analysis,
which is $\sim$80$\%$ of the total exposure. To create images, spectra, and light curves, we selected events with FLAG $=$ 0 and PATTERN $\leq$ 12 for MOS1/2 and PATTERN $\leq$ 4 for pn. 

\subsection {X-ray images}

Figure 1 (left panel) shows a color composite XMM-${\it Newton}$ image G296.8-0.3 in three energy bands: 0.5-1.0 (red), 1.0-2.0 (green), and 2.0-5.0 keV (blue), for the combined MOS1/2 cameras. Images were corrected for the spatial dependent exposure, and the instrumental background was also substracted. At X-ray energies $>$ 2.6 keV the SNR is not detected. The contour map overplotted on Figure 1 is the MOST 843 MHz radio continuum map (Whiteoak \& Green 1996), which can help us to identify the origin of the X-rays. The X-ray structure of the SNR is complex and covers $\sim$ 30\% of the total radio extent. It shows three different  components: interior diffuse emission coincident with the unusual rectangular strip (Gaensler et al. 1998) running through its center seen at radio frequencies (indicated with the dashed line in Fig.1, left panel), a bright soft shell-like feature with an angular size of $\sim$ 8', coincident with the northwest radio shell, and at least 9 point-like sources. All of them are numbered in Fig.1, right panel. These point sources were detected by using the source detection meta-task `edetect-chain'. Sources 1,2,3,4,5,6,7 and 9 display medium and hard X-ray emission and could be background AGNs. The remaining one, source 8, displays soft and medium X-ray emission. As can be seen, source 8 is located close to the geometrical center of the radio structure ($\alpha$=11$^h$58$^m$36.2, $\delta$=$-$62$^{\circ}$35$^\prime$20.0 ; J2000). The source is catalogued as 2XMMi J15836.1-623516 in the XMM-{\it Newton} Serendipitous Source Catalogue 2XMMi-DR3 (Watson et al. 2009)\footnote{http://vizier.u-strasbg.fr/viz-bin/VizieR?-source=IX}.

The interior of the radio remnant is filled with X-ray emitting material, and the spatial coincidence of X-ray and radio emission suggests that the physical conditions of the terminal shock region are very similar to those found at the outer shocks of ordinary SNRs. From the X-ray image we see enhanced X-ray emission on the northwest part of the SNR, which is probably caused by density enhancements in the medium in which the shock propagates and forms a more or less continuous structure. We notice that the X-ray emission is entirely contained within the boundaries of the radio shell. This bright filamentary X-ray structure is centered at ($\alpha$=11$^h$57$^m$44.5, $\delta$=$-$62$^{\circ}$33$^\prime$00.0; J2000) and has an angular size of 8 arcmin on the plane of the sky. The X-ray emission of the SNR fades out to the southeast, lacking a clearly defined edge.

\subsection{Spectral X-ray analysis}

\subsubsection{The diffuse X-ray emission}

In order to study the physical properties of the plasma in the remnant, a X-ray spectrum of G298.6-0.3 was extracted from the EPIC cameras using an elliptical region (shown in Fig. 1, right panel) that comprises $\sim$ 50$\%$ of the central CCD of the MOS cameras and spread part of 4 CCDs of the pn camera. For such purpose we use the SAS task `evselect' with suitable parameters for the MOS 1/2 and PN cameras. The background spectrum of the SNR was also taken from an elliptical region located within the central CCD of the MOS cameras. These regions were defined exclusively on the central chip to avoid having to account for chip-to-chip variations. The spectral analysis was performed with the XSPEC package (Arnaud, 1996). The X-ray spectrum of the SNR is shown in Fig. 2A (left panel).

The global spectrum has two components. The diffuse X-ray emission of the SNR (dominant in the energy range of 0.5$-$3.0 keV), plus the contribution of point-like sources (mainly contributing in energy range of 2.5$-$5.0 keV). In order to obtain the X-ray spectrum of the SNR, we have excluded all the point-like sources taking circular regions with a radius of 15$"$. The extracted EPIC MOS1/2 and pn spectra were grouped with a minimum of 30 counts per spectral bin, and the $\chi^{2}$ statistics was used. Ancillary response files (ARFs) and redistribution matrix files (RMFs) were calculated. The X-ray emission of the SNR peaks in the 0.5--3.0 keV energy range and it is clearly dominated by thermal emission. Thus, we used a PSHOCK model affected by an absorption interstellar model (PHABS; Balucinska-Church and McCammon 1992) to fit it. The X-ray parameters for the best-fit of the diffuse emission are given in Table 1.

\begin{table*}
\caption{X--ray spectral parameters for the diffuse X-ray emission.} 
\label{spec}
\begin{center}
\begin{tabular}{llc}
\hline
Model \& && Values \\
Parameters &      & \\
\hline
{\bf PHABS} && \\
$N_{\rm H}$ [cm$^{-2}$]& & 0.64($\pm$0.14)$\times$10$^{22}$  \\
\hline
{\bf PSHOCK}& &  \\
\hline
$kT$ [keV] &  &  0.86$\pm$ 0.1\\
Abund. [Z$\odot$]       &  &  0.12$\pm$ 0.08  \\
$\tau$[s cm$^{-3}$] & &  6.1 ($\pm$1.6)$\times$10$^{10}$ \\
Norm && 2.2($\pm$0.6)$\times$10$^{-3}$ \\
EM ($\times$10$^{57}$) &&  2.1 $\pm$0.2 \\
\hline
$\chi^{2}_{\nu}$ / d.o.f. && 1.34 / 160  \\
\hline
Flux(0.5-1.0)[erg~cm$^{-2}$~s$^{-1}$] & & 2.4($\pm1.5$)\,$\times$10$^{-12}$  \\
Flux(1.0-2.0)[erg~cm$^{-2}$~s$^{-1}$] & & 4.9($\pm1.2$)$\times$10$^{-13}$  \\
Flux(2.0-3.0)[erg~cm$^{-2}$~s$^{-1}$] & & 8.4($\pm1.8$)$\times$10$^{-14}$  \\
\hline
Total Flux(0.5-3.0)[erg~cm$^{-2}$~s$^{-1}$] & & 3.7($\pm1.7$)$\times$10$^{-12}$  \\
\hline
\hline
\end{tabular}
\end{center}
Normalization is defined as 10$^{-14}$/4$\pi$D$^2$$\times \int n_H\,n_e dV$,
where $D$ is distance in [cm], n$_{\sc H}$ is the hydrogen density [cm $^{-3}$], $n_e$
is the electron density [cm$^{-3}$], and $V$ is the volume [cm$^{3}$]. The flux 
in the three energy ranges 0.5-1.0, 1.0-2.0 and 2.0-3.0 keV, is absorption-corrected. 
Values in parentheses are the single parameter 90\% confidence interval.
The abundance parameter is given relative to the solar values of Anders \& Grevesse (1989).
\end{table*} 

\subsubsection{The X-ray source 2XMMi J115836.1$-$623516}

Since the point-like source is surrounded by diffuse X-ray emission of the SNR, we extracted its spectrum from a circular region with a radius of 6 arcsec (EE$\sim$ 50\%) and grouped it with a minimun of 18 counts per spectral bin. The background spectrum was estimated from an annular region with radii of 8 and 25 arcsec. The spectrum of the source is shown in Figure 2 (right panel). 

In order to study the X-ray properties of this compact source we fitted its spectrum with several spectral models. However, due to the low-photons statistic, the most representative one is a simple power-law (PL) model that yields a neutral hydrogen absorption column $N_{\rm H}$=0.55 $\pm$0.1 cm$^{-2}$, an index $\Gamma$=4.3 $\pm$0.7 and a normalization of 6.1($\pm$2.5)$\times$10$^{-6}$ cm$^{-2}$. The absorption corrected X-ray flux is F$_{\rm x}$=9.9 ($\pm$0.1) $\times$10$^{-14}$ erg s$^{-1}$ cm$^{-2}$ in the 0.3$-$3.0 keV band. The fit is acceptable in terms of the minimum $\chi^{2}$ ($\chi_{\nu}^{2}$= 1.1 for 24 d.o.f).

As can be seen, this model provides a good value of $\chi^{2}$ and the value of  $N_{\rm H}$ is similar to the value obtained for the SNR. This fact, supports the possibility that the point source detected at the geometrical centre of the SNR has a real physical connexion with G296.8-0.3. 



In addition, we found no significant pulsed signal with a period greater than twice the read-out time of the EPIC-PN camera in the
FF mode (73.3 ms), which corresponds to a Nyquist limit of 0.146 s.

\section {The Infrared emission from G296.8-0.3}

Using radio observations performed with the MOST radiotelescope at 0.843 GHz, and infrared Spitzer-MIPS (Rieke et al. 2004) observations of the SNR, we have investigated the positional correlation between all the detected emissions.
The MIPS basic calibrated data (BCD) were downloaded from the Spitzer archive.
These images were re-processed with the regular MIPS pipeline
(version S18.7.0), and then mosaicked using MOPEX (version
18.3.1) and the standard MIPS 24$\mu$m mosaic pipeline.

In Fig. 3, we show the MIPS image at 24 $\mu$m with the radio contours superimposed. The 24 $\mu$m emission is strongly correlated with the radio shells, with several faint filaments coincident on the northwest and southeast parts of the SNR. This emission is 
generally interpreted as thermal emission from dust grains that have been swept up and shock-heated by the supernova blast wave (Tappe et al. 2006).  The extraction regions used for computing the infrared fluxes on the shell-like boundaries are indicated in green. The mid-infrared fluxes at 24 $\mu$m of the northwest and southeast regions are 30.2$\pm$0.2 Jy and 14.3$\pm$0.1 Jy, respectively.

The X-ray emission detected on the northwest radio shell follows the infrared emission 
very well, which demonstrates the connection between the X-ray emitting plasma and the heated dust
grains. This result seems to indicate that the enhanced X-ray emission is caused by the expansion through a dense ISM 
with a density gradient toward the northwest side of the remnant. Under these conditions, it is possible to roughly make quantitative estimates of the swept-up ISM dust mass in the northwest and southeast rims using the formulae introduced by Whittet (2003),

\begin{equation}
\label{eqn;dustmass}
M_{{\rm{dust}}}  = \frac{{4\rho F_\lambda  d^2 }}{{3B_\lambda  (T_{{\rm{dust}}} )}}\left[ {\frac{a}{{Q_\lambda }}}
\right],
\end{equation}

In this equation we have adopted a density $\rm \rho=2500\,kg\,m^{-3}$ for silicate/graphite grains, a distance $d=9 \rm \,kpc$ for
the SNR, a dust temperature $T_{\rm dust}$ $\sim$ 100 K at 24$\mu$m, and an average ratio of grain radius $a$ over emissivity $Q_\lambda$ of $1\times10^{-5}\rm m$ (see, Tappe et al. 2006). Using the infrared flux densities $F_\lambda$ computed at 24$\mu$m and the Planck function $B_\lambda$($T_{dust}$) at 24 $\mu$m, we derived a dust mass of $\sim$ 0.010 $M_{{\rm{\odot}}}$ and $\sim$ 0.006  $M_{{\rm{\odot}}}$ for the northwest and southeast rims, respectively. Eq. 1 is an approximation assuming spherical dust grains of uniform size, composition, and in thermal equilibrium.
  
In order to inspect if there exists some infrared counterpart to 2XMMi J115836.1-623516, we show in Fig. 4 a \textit{Spitzer}/IRAC deep image of the central region of G296.9-0.3 in the 3.6~$\mu$m band. The extraction circle with a radius of 6 arcsec, used in the X-ray spectral analysis, is overplotted. As can be seen only one infrared source lies near the edge of the encircled region. The position of the X-ray source is given 
in the 2XMM catalogue (Watson et al., 2009) with a relative 
precision of one arcsec. However, it is well-known from 
cross-correlations with other astrometric catalogues that
the absolute accuracy in the position of \textit{XMM-Newton} sources
ranges between 5 and 10 arcsec (Della Ceca et 
al. 2004, L\'opez-Santiago et al. 2007, Combi et al. 2011). Therefore, we cannot assure if the 
infrared source observed near the edge of the encircle region is physically associated or not with the X-ray object.
 

\begin{figure}[t]
\includegraphics[width=8.4cm]{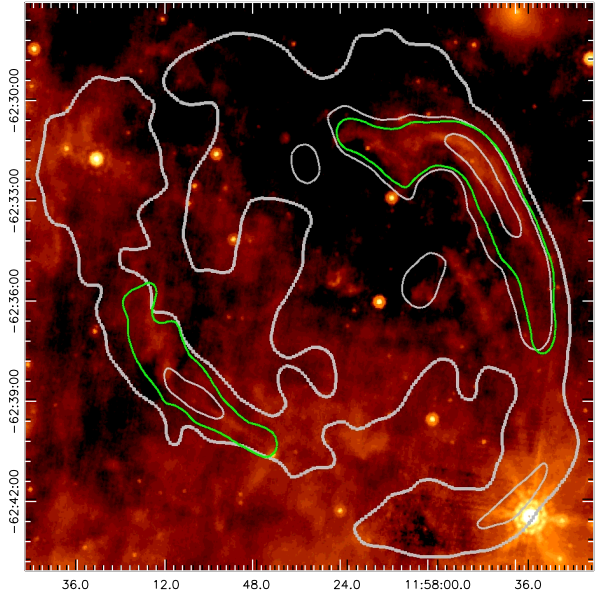}
   \caption{Spitzer MIPS 24 $\mu$m image of G296.8-0.3 with the radio contours at 843 MHz (in white) overlaid. The extraction regions used for computing the infrared fluxes are indicated in green color.}
    \end{figure}

\begin{figure}[t]
\includegraphics[width=8.0cm]{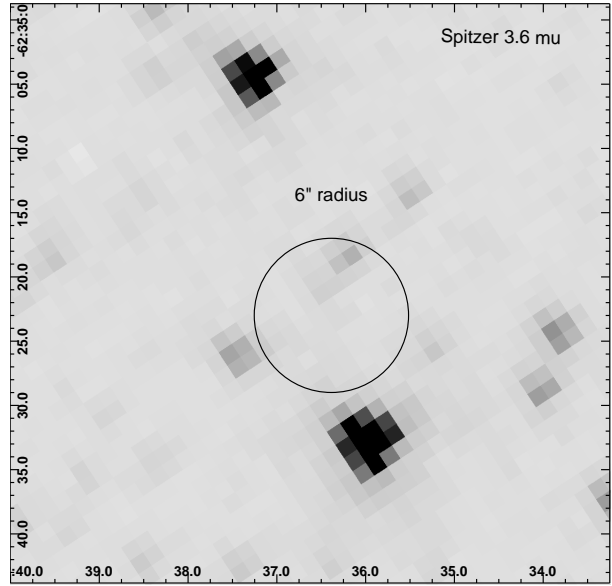}
\caption{Central region of the SNR observed by Sptizer in the 3.6 $\mu$m  band. We have overplotted the 6 arcsec extraction 
region used for the spectral X-ray analysis.}
\end{figure}

\section {Discussion}

\subsection{Physical parameters of G296.8-0.3}

Two possible evolutionary scenarios to explain the unusual morphology observed at radio frequencies 
in SNR G296.8$-$0.3, were initially studied by Gaensler et al. (1998). These authors suggested that 
the biannular appearance is induced by axial symmetry in the progenitor wind, or the SNR 
morphology resulted from the inhomogeneous ISM into which it is expanding. They conclude that the second possibility is more probable and that G296.8$-$0.3 seems consistent with being the remnant of a single explosion, where the unusual linear feature
running north-south through the remnant may represent a low density tunnel which has been re-energized by an encounter with
the SN shock.

It is clear that G296.8-0.3 is found in a complex area in the Scutum-Crux arm of our galaxy, where the ISM is particularly inhomogeneous and density variations in the pre-shock medium are present. The radio, infrared and X-ray observations of G296.8-0.3 here analyzed, can provide crucial information about the origin and evolution of the SNR, as well as regarding the age, energetics, ambient conditions and the presence of heated dust. With all this information in mind, we could outline a possible framework that allows to explain the characteristics of the emissions observed from the SNR. 

The XMM observations reveal that there is diffuse X-ray emission in the interior of the SNR well-correlated with the unusual rectangular strip running through its center seen at radio frequencies, a bright soft shell-like feature coincident with the internal northwest radio shell, and several hard point-like sources (possibly background AGNs). Moreover, the 24 $\mu$m observations show two limb-brightened shell-like structures on the northwest and southeast parts of the SNR, and faint filaments strongly correlated with the radio shells. 

The emission measure (EM) computed for the global region of the SNR can allow us to estimate the corresponding 
density of the X-ray emitting gas. From the X-ray image, we can roughly assume that the X-ray emission fills an ellipsoid with radii of $\sim$ 3 $\times$ 4 arcmin and estimate the volume $V$ of the X-ray emitting plasma. At a distance of 9 kpc, the SNR defines an X-ray emitting volume $V_{\rm SNR}$= 6.1$\times$10$^{58}$ cm$^{3}$. Based on the EM determined by the spectral fitting (see Table 1), we can estimate the electron density of the plasma, $n_{e}$, by $n_{e}$=$\sqrt{EM/V}$, which results in $n_{e}$$\sim$ 0.18 cm$^{-3}$. In this case, the number density of the nucleons was simply assumed to be the same as that of electrons. The age $t$ is determined from the ionization timescale, $\tau$, by $t$=$\tau$/$n_{e}$. Therefore, the elapsed time after the plasma was heated is $t$ $\sim$ 1.0$\times$10$^{4}$ yr. This result shows that G296.8$-$0.3 is a middle-aged SNR.


Assuming that the SNR is in the adiabatic (Sedov-Taylor) phase, Gaensler et al. (1998) obtained several physical parameters for the object.  Taking into account a kinetic energy of the initial explosion of $E$= 10$^{51}$ erg, and that the SNR expands in an ISM with a density of 0.2 cm$^{-3}$, these authors computed an age of $t$=(10$\pm$2)$\times$10$^{3}$ yr for G296.8-0.3. This value agrees very well with the age obtained by us above using the X-ray information.



\subsection{The central source 2XMMi J115836.1-623516}

Concerning the point-like X-ray source located close to the geometrical center of the radio structure, we see no significant variability that disfavors an accreting binary origin, a soft thermal spectrum that eliminates a background active nucleus, lack of radio counterpart, and absence of a surrounding pulsar wind nebula. At first sight, we can see that the source displays some characteristics of the so-called CCO (see Pavlov et al. 2004, for a review), a new population of isolated neutron stars (NSs) with clear differences from isolated rotation-powered pulsars and accretion-powered X-ray pulsars in close binary systems.


The nature of these objects is still unclear. It is thought that the X-ray emission from CCOs is generally due to the thermal cooling of the NS (e.g Zavlin, Trumper \& Pavlov 1999), with typical temperatures of a few 10$^{6}$ K, as inferred from their thermal-like spectra. They have X-ray luminosities ($L_{\rm X}$) in the range of 10$^{33}$-10$^{34}$ erg s$^{-1}$ and display X-ray spectra characterized by a blackbody model with temperatures ($kT$) in the range of 0.2-0.5 keV or a power-law model with very steep index $\Gamma$ (see Pavlov et al. 2003). Halpern \& Gotthelf (2010) have recently suggested that these objects could be weakly magnetized NSs ($B\sim10^{10}$ G), i.e., a kind of ``anti-magnetars''. 

In order to check the characteristics of 2XMMi J115836.1-623516, we computed its $L_{\rm X}$ and spin-down luminosity $\dot{E}$, to compare with other well-known CCOs (Pavlov et al. 2003). Adopting a mean distance of 9 kpc and a total unabsorbed X-ray flux of  $F_{0.3-3.0}$=9.9$\times$10$^{-14}$ ergs cm$^{-2}$ s$^{-1}$, we obtain an unabsorbed luminosity $L_{\rm X}$= 1.0$\times$10$^{33}$ ergs s$^{-1}$. A rough estimate of the spin-down luminosity can be derived using the empirical formula by Seward \& Wang (1988), log $L_{\rm X}$ (ergs s$^{-1}$)= 1.39 log $\dot{E}$ - 16.6, which gives $\dot{E}$= 3.9$\times$10$^{35}$ ergs s$^{-1}$. $L_{\rm X}$ lies within the range suggested by Pavlov et al. (2004) and by Halpern \& Gotthelf (2010) for CCO objects. The second quantity, $\dot{E}$, falls below the empirical threshold for generating bright wind nebulae of $\dot{E_{c}}$ $\approx$ 4$\times$10$^{36}$ ergs s$^{-1}$. These results suggest that the system G296.8-0.3/2XMMi J115836.1-623516 is a thermal SNR with, possibly, a nondetected NS. Plausible reasons for the nondetection of a NS are the low-photon statistic, a short rotation period or unfavorable geometrical conditions.

\section{Conclusions}

We have analyzed radio, infrared and X-ray data of the SNR G296.8-0.3 to investigate the origin of the radiative process involved in the generation of the emission observed. 
The diffuse X-ray emission is clearly correlated with the unusual rectangular strip running through its center seen at radio frequencies, where we found a region with significant lower density, and a bright X-ray shell-like feature at the northwest part of the SNR coincident with the internal boundary of the radio shell. The 24 $\mu$m observations show two limb-brightened shell-like structures on the northwest and southeast parts of the SNR, and faint filaments strongly correlated with the radio shells. The spectral study confirms that the X-ray diffuse emission is thermal and the column density of the SNR is high ($N_{\rm H}$$\sim$0.64$\times$10$^{22}$ cm$^{-2}$) supporting distant location ($d >$9 kpc) for the SNR. 

In addition, a compact X-ray source was also detected close to the geometrical center of the SNR. 
The object presents some characteristics of the CCOs, and the neutral hydrogen absorption column $N_{\rm H}$ is consistent with that of the SNR. Although these results support a physical connexion with the SNR, high-resolution X-ray observations carry out with the Chandra satellite are necessary to better understand the nature of the X-ray source.

\acknowledgments

The authors acknowledge support by DGI of the Spanish Ministerio de Educaci\'on y Ciencia under grants AYA2010-21782-C03-03, FEDER funds, Plan Andaluz de Investigaci\'on Desarrollo e Innovaci\'on (PAIDI) of Junta de Andaluc\'{\i}a as research group FQM-322 and the excellence fund FQM-5418. J.A.C. and J.F.A.C. are researchers of CONICET. J.F.A.C was supported by grant PICT 2007-02177 (SecyT). J.A.C was supported by grant PICT 07-00848 BID 1728/OC-AR (ANPCyT) and PIP 2010-0078 (CONICET). J.L.S. acknowledges support by the Spanish Ministerio de Innovaci\'on y Tecnolog\'ia under grant AYA2008-06423-C03-03.


\begin{thebibliography}{}

\bibitem[Anders 
\& Grevesse(1989)]{1989GeCoA..53..197A} Anders, E., \& Grevesse, N.\ 1989, \gca, 53, 197 

\bibitem[Arnaud(1996)]{1996ASPC..101...17A} Arnaud, K.~A.\ 1996, 
Astronomical Data Analysis Software and Systems V, 101, 17

\bibitem[Aschenbach(1998)]{1998PhST...77..122A} Aschenbach, B.\ 1998, Physica Scripta Volume T, 77, 122

\bibitem[Balucinska-Church 
\& McCammon(1992)]{1992ApJ...400..699B} Balucinska-Church, M., \& McCammon, D.\ 1992, \apj, 400, 699 

\bibitem[Bamba et al.(2003)]{2003ApJ...589..827B} Bamba, A., Yamazaki, R., 
Ueno, M., \& Koyama, K.\ 2003, \apj, 589, 827

\bibitem[Bocchino et al.(2005)]{2005A&A...442..539B} Bocchino, F., van der Swaluw, E., Chevalier, R., \& Bandiera, R.\ 2005, \aap, 442, 539 

\bibitem[Combi et al.(2010)]{2010A&A...522A..50C} Combi, J.~A., et al.\ 2010a, \aap, 522, A50

\bibitem[Combi et al.(2010)]{2010A&A...523A..76C} Combi, J.~A., et al.\ 2010b, \aap, 523, A76 


\bibitem[Combi et al.(2011)]{2011Ap&SS.331...53C} Combi, J.~A., et al.\ 2011, \apss, 331, 53 

\bibitem[Della Ceca et al.(2004)]{2004A&A...428..383D} Della Ceca, R., et al.\ 2004, \aap, 428, 383 




\bibitem[Gaensler et al.(1997)]{1997ApJ...479..845G} Gaensler, B.~M., 
Manchester, R.~N., Staveley-Smith, et al. 1997, \apj, 479, 845


\bibitem[Gaensler et al.(1998)]{1998MNRAS.296..813G} Gaensler, B.~M., 
Manchester, R.~N., \& Green, A.~J.\ 1998, \mnras, 296, 813

\bibitem[Goss 
\& Shaver(1970)]{1970AuJPA..14....1G} Goss, W.~M., \& Shaver, P.~A.\ 1970, Australian Journal of Physics Astrophysical Supplement, 14, 1 


\bibitem[Halpern 
\& Gotthelf(2010)]{2010ApJ...709..436H} Halpern, J.~P., \& Gotthelf, E.~V.\ 2010, \apj, 709, 436 

\bibitem[Hwang 
\& Markert(1994)]{1994ApJ...431..819H} Hwang, U., \& Markert, T.~H.\ 1994, \apj, 431, 819 

\bibitem[Large 
\& Vaughan(1972)]{1972Natur.236..117L} Large, M.~I., \& , A.~E.\ 1972, \nat, 236, 117 

\bibitem[L{\'o}pez-Santiago et al.(2007)]{2007A&A...463..165L} L{\'o}pez-Santiago, J., Micela, G., Sciortino, S., Favata, F., Caccianiga, A., Della Ceca, R., Severgnini, P., \& Braito, V.\ 2007, \aap, 463, 165 


\bibitem[Pavlov et al.(2003)]{2003ApJ...591.1157P} Pavlov, G.~G., Teter, 
M.~A., Kargaltsev, O., \& Sanwal, D.\ 2003, \apj, 591, 1157

\bibitem[Pavlov et al.(2004)]{2004IAUS..218..239P} Pavlov, G.~G., Sanwal, 
D., \& Teter, M.~A.\ 2004, Young Neutron Stars and Their Environments, 218, 239 


\bibitem[Rieke et al.(2004)]{2004ApJS..154...25R} Rieke, G.~H., et al.\ 
2004, \apjs, 154, 25

\bibitem[Sasaki et al.(2004)]{2004ApJ...617..322S} Sasaki, M., Plucinsky, P.~P., Gaetz, T.~J., Smith, R.~K., Edgar, R.~J., \& Slane, P.~O.\ 2004, \apj, 617, 322 

\bibitem[Schwentker(1994)]{1994A&A...286L..47S} Schwentker, O.\ 1994, \aap, 286, L47

\bibitem[Shaver 
\& Goss(1970)]{1970AuJPA..14...77S} Shaver, P.~A., \& Goss, W.~M.\ 1970, Australian Journal of Physics Astrophysical Supplement, 14, 77 


\bibitem[Seward \& Wang(1988)]{1988ApJ...332..199S} Seward, F.~D., \& Wang, Z.-R.\ 1988, \apj, 332, 199 


\bibitem[Str{\"u}der et 
al.(2001)]{2001A&A...365L..18S} Str{\"u}der, L., et al.\ 2001, \aap, 365, L18

\bibitem[Tappe et al.(2006)]{2006ApJ...653..267T} Tappe, A., Rho, J., 
\& Reach, W.~T.\ 2006, \apj, 653, 267

\bibitem[Turner et al.(2001)]{2001A&A...365L..27T} Turner, M.~J.~L., et al.\ 2001, \aap, 365, L27 


\bibitem[Watson et al.(2009)]{2009A&A...493..339W} Watson, M.~G., et al.\ 2009, \aap, 493, 339 

\bibitem[Whiteoak \& Green(1996)]{1996A&AS..118..329W} Whiteoak, J.~B.~Z., \& Green, A.~J.\ 1996, \aaps, 118, 329 

\bibitem[Whittet 2003]{2003IOP..2nd} Whittet, D. C. B. 2003, Dust in the Galactic Environment (2nd ed.; Bristol: IOP)

\bibitem[Yamaguchi et al.(2004)]{2004PASJ...56.1059Y} Yamaguchi, H., Ueno, 
M., Koyama, K., Bamba, A., \& Yamauchi, S.\ 2004, \pasj, 56, 1059

\bibitem[Zavlin et al.(1999)]{1999ApJ...525..959Z} Zavlin, V.~E., Tr{\"u}mper, J., \& Pavlov, G.~G.\ 1999, \apj, 525, 959  


\end{thebibliography}
\end{document}